\documentclass[twocolumn,floatfix,aps,prl,showpacs]{revtex4-1}
\usepackage{graphicx}

\begin{document}
\title{Experimental verification of democratic particle motions
by direct imaging of glassy colloidal systems}
\author{J. A. Rodriguez Fris$^\dag$}
\author{G. A. Appignanesi$^\dag$}
\author{Eric R. Weeks$^\ddag$}
 \affiliation{
$^\dag$Area Fisicoqu\'{\i}mica, INQUISUR-UNS-CONICET and Departamento de Qu\'{\i}mica, Universidad Nacional del Sur, Av. Alem 1253, 8000 Bah\'{\i}a Blanca, Argentina.\\
$^\ddag$Physics Department, Emory University, Atlanta, GA 30322\\}

\date{\today}


\begin{abstract}
We analyze data from confocal microscopy experiments of a
colloidal suspension to validate predictions of rapid sporadic
events responsible for structural relaxation in a glassy sample.
The trajectories of several thousand colloidal particles are
analyzed, confirming the existence of rapid sporadic events
responsible for the structural relaxation of significant regions
of the sample, and complementing prior observations of dynamical
heterogeneity.  The emergence of relatively compact clusters
of mobility allows the dynamics to transition between the large
periods of local confinement within its potential energy surface,
in good agreement with the picture envisioned long ago by Adam
and Gibbs and Goldstein.
\end{abstract}


\maketitle


A complete understanding of the molecular underpinnings
of glassy relaxation (the dramatic dynamical slowing
down that arises when a liquid is rapidly cooled
below its melting point avoiding crystallization),
remains a major challenge in condensed matter physics
\cite{angell_91,ediger_00,debenedetti_01,ediger_96,gotze_99,proceedings_02}.
As long ago as 1965, Adam and Gibbs \cite{adam_65}
proposed an appealing picture that accounted for this
enormous increase in relaxation timescales within a narrow
temperature window.  They suggested that the dynamics of
a glass-forming supercooled liquid proceeds by means of
cooperatively rearranging regions (CRR) whose size and relaxation
timescale grow considerably as temperature is lowered, giving
the decrease in configurational entropy of the system 
\cite{angell_91,ediger_00,debenedetti_01,ediger_96,gotze_99,proceedings_02,adam_65}.
This description suggests that a supercooled liquid at low
temperatures can be decomposed in independently relaxing
compact subsystems (the CRR) whose molecules attempt to
change configuration, but which can only undergo a transition when
they rearrange in a concerted manner.  Thus each of them
(in the words of Adam and Gibbs \cite{adam_65}) surmounts,
essentially simultaneously, the individual barrier restricting its
arrangement.  In this picture, it is expected that each region of a supercooled
liquid should be practically ``frozen'' in a given portion
of configuration space for large times (larger as temperature
decreases given the growing size of the regions and thus of the
number of molecules involved in the rearrangement) and then will relax
(asynchronously and independently of other regions) by having
a burst of mobility characterized by the sharp emergence of a
compact cluster of mobile particles\cite{adam_65,goldstein}.
Hence, at any given time the system would present dynamics
that would vary significantly from one region to another:
the dynamics should then be heterogeneous {\em in space}
\cite{angell_91,ediger_00,debenedetti_01,ediger_96,gotze_99,proceedings_02}.

The validity of such a heterogeneous scenario has been confirmed
both experimentally and computationally, since the existence of
dynamical heterogeneities
\cite{schmidt_91,donati_98,richert_02,kegel_00,weeks_00,weeks_02,kob_97,butler_91,cicerone_95}
has been detected. Simulations of model glassy systems have shown
that the more mobile particles are not homogeneously distributed
in space but arranged in (non-compact) clusters \cite{donati_98}.
The time scale for the motion of these more mobile particles is
$t^*$, a timescale close to the structural $\alpha$-relaxation
time, $\tau_\alpha$; $\tau_\alpha$ is calculated as the time
scale when the incoherent self intermediate scattering function
has decayed to 1/e.  These results have also received
experimental support in colloidal suspensions (experimental
models for glassy relaxation) \cite{kegel_00,weeks_00}.

More recently \cite{appignanesi_06}, computational studies have determined that within any dynamically heterogeneous region of the system, the relaxation is not gradual but also heterogeneous {\em in time}, since the $\alpha$ relaxation is almost exclusively governed by rapid sporadic events characterized by the emergence of relatively compact clusters of mobile particles (termed as ``democratic'' clusters or d-clusters \cite{appignanesi_06}). These events trigger transitions between local metabasins (MB, basins of attraction of the potential energy surface comprising a group of similar closely-related structures or local minima \cite{appignanesi_06,debenedetti_01, doliwa_03,vogel_04} where the system is confined for long times). These cooperatively relaxing units or d-clusters have been identified in molecular dynamics simulations of different glassy systems like a binary Lennard-Jones \cite{appignanesi_06} system, supercooled water \cite{agua_07} and amorphous silica \cite{ariel-thesis} and represent natural candidates for the CRR proposed by Adam and Gibbs \cite{adam_65}. A recent inhomogeneous mode-coupling theory of dynamical heterogeneity has related them to the (fractal) geometrical structures carrying the dynamical correlations at timescales commensurable with that of the $\alpha$ relaxation (more compact than the open-like structures expected at much shorter timescales) \cite{biroli_06}. Additionally, a recent experimental and computational work in a glassy polymer provided indirect experimental support to the MB-MB transitions and d-clusters \cite{vallee_07}. However, experiments with molecular glasses lack the level of resolution necessary to directly observe them and thus, no direct experimental information has verified the existence of such events up to date. 

In this work we study a colloidal suspension (an excellent experimental model of glassy systems with particles big enough to be directly observed by confocal microscopy) and we shall demonstrate that detailed tracking of particle motions is indeed able to detect the aforementioned kind of events.

We analyze the data of Refs.~\cite{weeks_00,weeks_02}, taken from
confocal microscopy experiments of colloidal samples.  The
colloids are sterically stabilized colloidal
poly-(methylmethacrylate) with diameter $d=2.36$~$\mu$m and a
polydispersity of $\sim 5\%$.  They are dyed with rhodamine and
suspended in a density-matching and index-matching solvent
mixture of cycloheptylbromide and decalin.  In this solvent, the
colloidal particles possess a slight charge, and exhibit a glass
transition at a volume fraction $\phi \approx 0.58$.  A confocal
microscope rapidly acquires three-dimensional images once every
$10-20$~s.  The images are post-processed to locate particle
centers with an accuracy of $0.03$~$\mu$m in $x$ and $y$ and
$0.05$~$\mu$m in $z$.  Due to the difficulty of identifying
particles near the edges of the images, the useful data are
within a region of size $L_x=67$~$\mu$m, $L_y=62$~$\mu$m, and
$L_z=9$~$\mu$m, corresponding to several thousand particles.  The
volume fractions are determined by counting the particles within
a subvolume, and are known to within $\pm 0.01$ with the
uncertainty mainly due to the uncertainty of the particle
diameter ($\pm 0.01$~$\mu$m).  For further experimental details,
see Refs.~\cite{weeks_00,weeks_02}.

\begin{figure}[h!]
\includegraphics[width=0.9\linewidth]{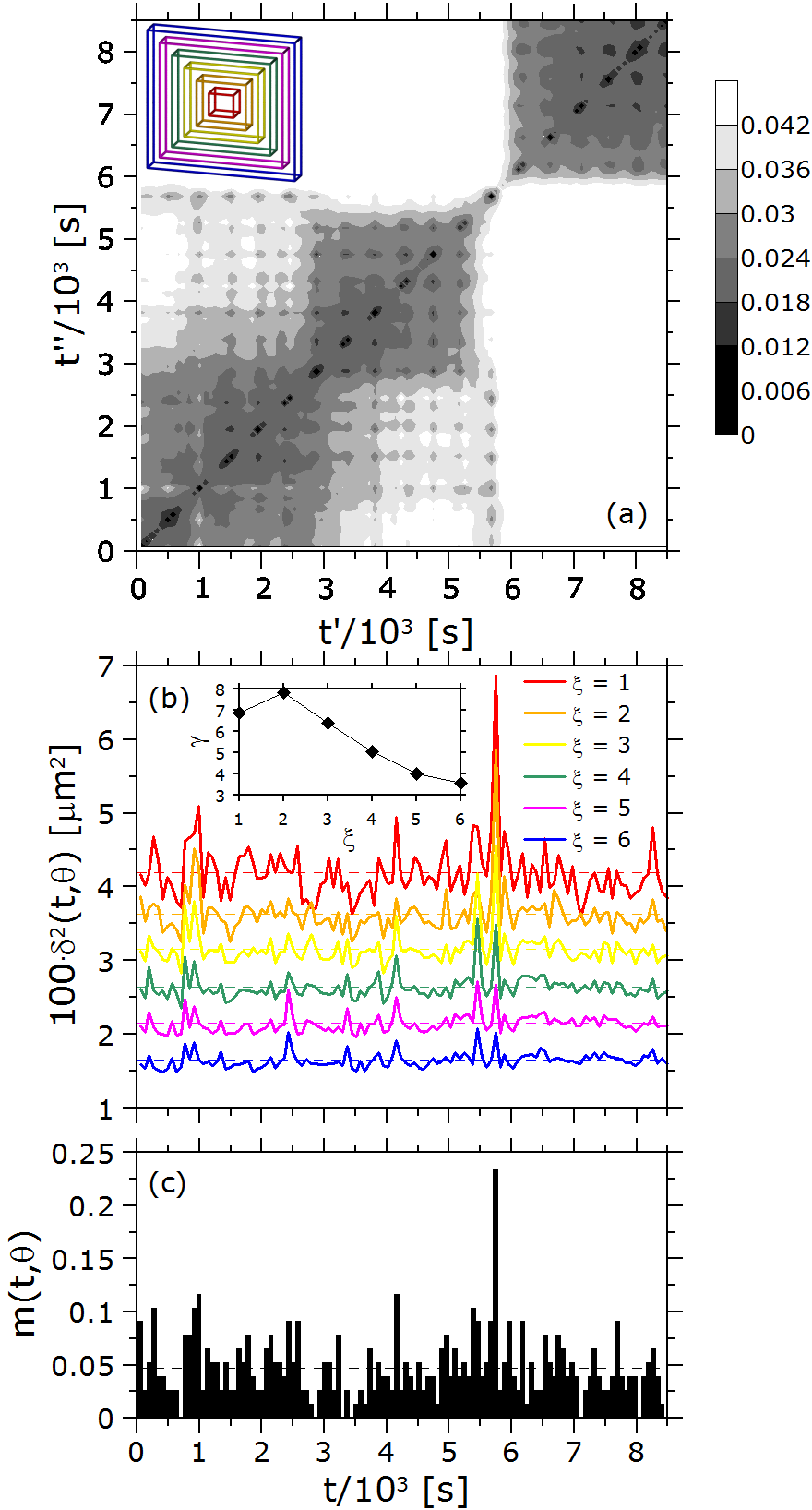}
\caption{
(a) Distance matrix $\Delta^2(t',t'')$ for portion $\xi=1$. The gray level corresponds to values of $\Delta^2(t',t'')$ that are given to the right of the figure. Units are $\mu$m$^2$. Inset: for the analysis, we followed the colloidal particles that were initially ($t=0$) within the boundaries of rectangular prisms: $\xi=6$ (blue, the complete system), magenta, green, yellow, orange and $\xi=1$ (red, smallest region) respectively. All $\xi$ have the same depth $L_z$. Colloidal particles are not shown. (b) Averaged squared displacement $\delta^2(t,\theta)$ for all $\xi$. Each series (analysis over each $\xi$) is shifted by 0.5 $\mu$m$^2$ respect the former one. For comparison we included the corresponding average values of $\delta^2(t,\theta)$ over all times, $\langle \delta^2(t,\theta) \rangle$ (dashed colored lines, also shifted). The value of $\theta$ is 72 s. Inset: $\gamma=|\delta^2_\xi(5706\,\,{\rm s},\theta)-\langle \delta^2(t,\theta) \rangle_\xi|/\sigma_\xi$ {\it vs} $\xi$, where $\sigma$ is the standard deviation in $\delta^2$ and 5706~s is the time of the largest average squared displacement. Subscript $\xi$ means that the function was evaluated for molecules in $\xi$. A maximum for $\xi=2$ is observed. (c) The function $m(t,\theta)$ for $\xi=1$ and its average value over all times (dashed black line). }
\label{fig1}
\end{figure}

Since at any given time a large system would consist of several
different CRR, we divided the experimental system into 6
subsystems or portions $\xi$, each one with an increasing number
of colloidal particles $N(\xi)$ (see the inset in
Fig.~\ref{fig1}(a)). All portions have the same depth $L_z$ and
center $(L_x/2,L_y/2,L_z/2)$. Portion $\xi$ comprises the
particles that were initially ($t=0$) \cite{ini} within the
boundaries of the corresponding rectangular prism of length $L_x
\cdot (\xi/6)$ and height $L_y \cdot (\xi/6)$.  For $\phi=0.56$, the number of particles within each portion is: $N(\xi \!= \!1)=77$, $N(2)=310$, $N(3)=703$, $N(4)=1255$, $N(5)=1962$ and $N(6)=2759$ particles. We shall present results for $\phi=0.56$ unless otherwise indicated, but similar results were obtained for $\phi=0.53$, $\phi=0.52$ and $\phi=0.46$.

To identify MBs we employed the following ``distance matrix'' ($\Delta^2$) function \cite{ohmine_95}:

\begin{equation}
\Delta^2(t',t'') = \frac{1}{N}\sum_{i=1}^N |{\bf r}_i(t')-{\bf r}_i(t'')|^2
\quad ,
\label{eq1}
\end{equation}

\noindent
where ${\bf r}_i(t)$ is the position of particle $i$ at time $t$. $\Delta^2(t',t'')$ gives the system normalized squared displacement in the time interval ($t'$, $t''$). 


\begin{figure}[tb]
\includegraphics[width=0.9\linewidth]{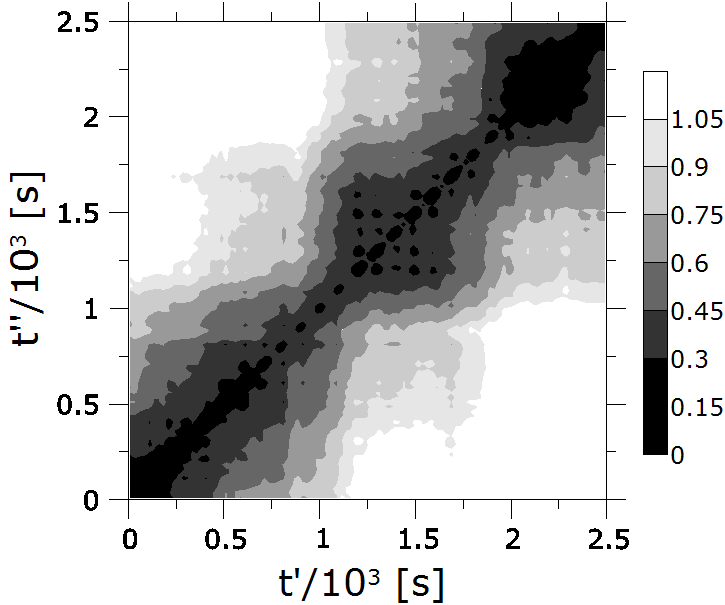}
\caption{
Similar to Fig.~\ref{fig1}(a), but for $\phi=0.46$ and for a
(rectangular prism) region of approximately the size of $\xi=1$.
At this $\phi$, $N(1)=44$ and
$t^*=300$~s.}

\label{fig2}
\end{figure}

A plot of $\Delta^2$ as a function of $t'$ and $t''$ can be seen
in Fig.~\ref{fig1}(a) for $\phi=0.56$. These results are typical
for all studied $\phi$ (as an example, in Fig.~\ref{fig2} we
show an equivalent plot for $\phi=0.46$). The darker the shading,
the smaller the distance between the configurations at times $t'$
and $t''$. From this figure we can learn that the dynamics of this
portion is quite heterogeneous {\it in time} in that it stays for
a significant time relatively close to one region in configuration
space, dark square-like regions, before it finds a pathway to
a new region. The value of $\Delta^2(t',t'')$ {\it within} a MB
is around 0.02~$\mu$m$^2$ as compared to values much larger than
0.04~$\mu$m$^2$ if the system is in {\it different} MBs (see legend
on the right of the figure).  If there
were no MBs, the plot would show a dark shadow at the diagonal
$t''=t'$ and a gradual decrease in shading perpendicular to it,
as it would be seen at low $\phi$ (similarly to the case of structural glasses at high temperature \cite{appignanesi_06})
and/or large systems;
compare Fig.~\ref{fig1}(a) ($\phi=0.56$) to Fig.~\ref{fig2}
($\phi=0.46$). These figures demonstrate that the system 
spends large amounts of time exploring the local MB, and only
occasionally moves on to a neighboring MB.
Indeed, we can see that this
trajectory resides within a MB for times much larger than $t^*
\approx 1000$~s, the maximum in the non-Gaussian parameter
$\alpha_2(t)$ \cite{kob_97,weeks_00}. We also point out that from
Fig.~\ref{fig1}(a) it is evident that the time for a MB-MB
transition is quite short, on the order of 70~s, which thus
corresponds to about 7\% of $t^*$.  In the lower volume fraction
data of Fig.~\ref{fig2}, the transitions are also
rapid although slightly less distinct.

\begin{figure}[tb]
\includegraphics[width=0.9\linewidth]{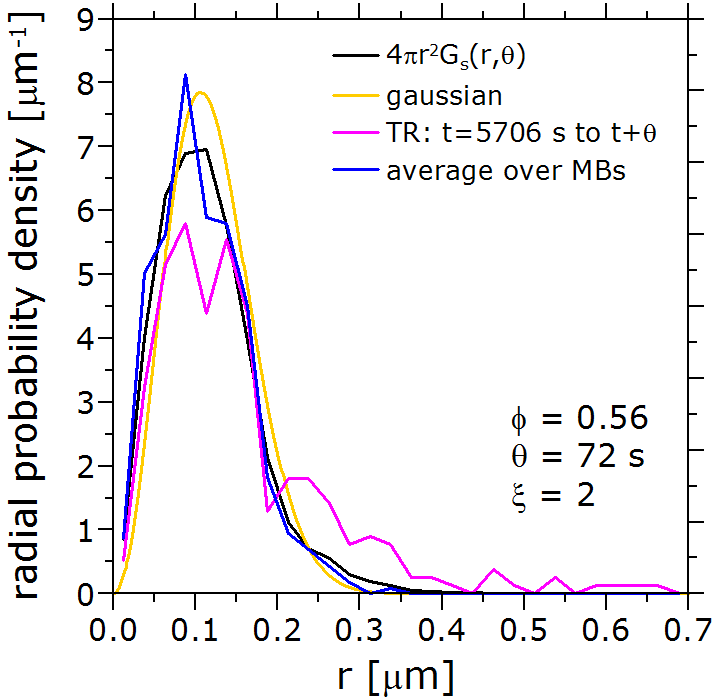}
\caption{
Distribution functions for $\xi=2$ and $\theta=72$~s. The black curve is $4 \pi
r^2 G_{\rm s}(r,\theta)$, the self-part of the van Hove function,
and the gold curve is a Gaussian with the same value of
$\langle r^2(\theta)\rangle$.  The crossing point of these two
curves at 
$r \approx 0.23$~$\mu$m is used as a threshold to identify the
democratically moving
particles. The blue and magenta curves are the average of $4
\pi r^2 \widehat{G}_{\rm s}(r,t,t+\theta)$ for different values
of $t$ in which the system is inside a MB (see text for details),
and a transition (TR) from $t=5706$~s to $t+\theta$.
}

\label{fig3}
\end{figure}

In Fig.~\ref{fig1}(b) we also show, for all $\xi$ and same time interval, $\delta^2(t,\theta)$, the averaged squared displacement of the particles within a time interval $\theta$ (solid curves). This function is defined as
\begin{eqnarray}
\delta^2(t,\theta)  & = & \Delta^2(t,t+\theta) \\
 & = & 
\frac{1}{N}\sum_{i=1}^N |{\bf r}_i(t)-{\bf r}_i(t+\theta)|^2\quad .
\label{eq2}
\end{eqnarray}
\noindent

A comparison of $\delta^2$ with the distance matrix shows that $\delta^2$ is showing pronounced peaks exactly when the system leaves a MB. Thus we see that changing the MB is indeed associated with a rapid significant particle motion as measured by $\delta^2$. Also included are the average values of $\delta^2(t,\theta)$ for all $\xi$ over all times, $\langle \delta^2(t,\theta) \rangle$, represented by dashed lines. It is clear that the larger systems have a lower relation between fluctuations in $\delta^2$ at the MB transitions and its average value: This ratio in Fig.~\ref{fig1}(b) for the peak at time $t=5706$~s is maximum at $\xi=2$ (see inset), thus providing an indication of the size of the MB-MB transition event. 

To understand the motion of the particles when the system leaves a MB we have calculated the function $4 \pi r^2 \widehat{G}_{\rm s}(r,t,t+\theta)$,
the distribution of displacement $r$ of the particles for a given time difference $\theta=72$~s. (Note that the average of $4 \pi r^2 \widehat{G}_{\rm s}(r,t,t+\theta)$ over $t$ gives $4 \pi r^2 G_{\rm s}(r,\theta)$, the self-part of the van Hove function). An average of this distribution is shown in Fig. \ref{fig3} (blue curve) for different values $t$ within a MB [$t$/s = 378, 2898, 3474, 4626, 6210 and 7506 for the case of Fig.~\ref{fig1}(a)]. Also included is the self-part of the van Hove function (black curve) and we can see that both curves are (within the noise of the data) identical, thus showing that in a MB the system moves basically the same as on average. We also show the distribution (magenta curve) for $t=5706$~s in which the system is about to leave a MB. For this value of $t$ the distribution is clearly displaced to the right with respect to $4 \pi r^2 G_{\rm s}(r,\theta)$, showing that in this time regime the motion of the system is much faster than on average. Thus we can conclude that the  peaks of the $\delta^2$ are {\it not} due to the presence of a {\it few} fast moving particles, but instead to a ``democratic'' movement of {\it many} particles.

\begin{figure}[tb]
\includegraphics[width=0.4\linewidth]{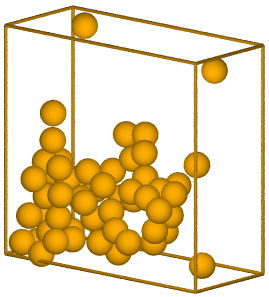}
\includegraphics[width=0.4\linewidth]{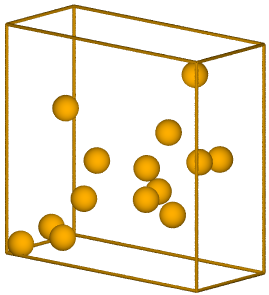}
\caption{
Actual size and position of democratic particles for: (left) a MB-MB
transition from $t = 5706$~s to $t+\theta$ and (right) $t=7290$~s
to $t+\theta$ inside a MB. The data are for $\xi=2$,
$\theta=72$~s, $\phi=0.56$.
}
\label{fig4}
\end{figure}

To further demonstrate that this is indeed the case and to explore the spatial distribution of mobility, we have defined as ``democratic'' all those particles that in the time interval $\theta=72$~s have moved more than $r_{\rm th}=0.23$~$\mu$m (a value very close to 0.25~$\mu$m, the size of the cage formed by particles surrounding a single particle \cite{weeks_02}), and denote the fraction of such particles by $m(t,\theta)$. We take the value of $r_{\rm th}$ as the second intersection between $4 \pi r^2 G_{\rm s}(r,\theta)$ and a Gaussian (see Fig.~\ref{fig4}, gold curve) with the same value of mean squared displacement, $MSD= \langle r^2(\theta)\rangle$ (however, other threshold choices yield similar results). In Fig.~\ref{fig1}(c) we have included the fraction $m$ of democratic particles for $\xi=1$ as a function of time (vertical bars). A comparison of this data with $\delta^2$ shows that $m$ is indeed large whenever $\delta^2$ increases rapidly. This fraction is on the order of 20 \% of the particles and thus significantly larger than one would expect from $4 \pi G_{\rm s}(r,\theta)$ if one integrates this distribution from $r_{\rm th}$ to infinity and which gives 0.051.
In turn, Fig.~\ref{fig4} (left) shows the 3D location of the democratic particles involved in a typical MB-MB transition for $\xi=2$ (we show the case for the event at $t=5706$~s in Fig.~\ref{fig1} but other cases display similar results). We can see that the particles are not homogeneously distributed in space (as is indeed the case in Fig.~\ref{fig4} (right) for $t=7290$~s inside a MB) but arranged in a relatively compact cluster.

In summary, the experimental results shown in this work for a
colloidal system provide direct validation to the picture of glassy
relaxation previously shown by MD simulations of several glassy systems
\cite{appignanesi_06,agua_07,ariel-thesis,vallee_07}: The dynamics
spends large times confined within a metabasin, interspersed with rapid bursts
in mobility characterized by the emergence of relatively compact
clusters of democratic particles which trigger the structural
or $\alpha$ relaxation.  Systems closer to the glass transition
have more distinct metabasin transitions that occur more
infrequently (compare
Fig.~\ref{fig1}(a) to Fig.~\ref{fig2}).
Thus, this behavior conforms to the
scenario put forth long time ago by Adam and Gibbs\cite{adam_65}
and Goldstein\cite{goldstein}.

Financial support from ANPCyT, SeCyT and CONICET is gratefully
acknowledged. GAA and JARF are research fellows of CONICET.  The
work of ERW was supported by a grant from the NSF (CHE-0910707).


\begin{thebibliography}{10}
\bibitem{ini}
$\langle r^2(\Delta t = 8000$~s$) \rangle \approx 0.11$ $\mu$m$^2$ for the main system, so we expect most of the particles to remain close of their initial position.
\end{thebibliography}
\end{document}